\documentclass[aps,prl,twocolumn,groupedaddress,showpacs]{revtex4}
\usepackage[dvips]{graphicx}
\usepackage{color}
\usepackage{natbib}
\usepackage{amsmath}

\begin{document}

\title{Effect of Intra-molecular Disorder and Inter-molecular Electronic
Interactions on the Electronic Structure of Poly-{\it p}-Phenylene
Vinylene (PPV)}

\author{Ping Yang, Enrique R. Batista, Sergei Tretiak, Avadh Saxena,
 Richard L. Martin, and D. L. Smith}

\email[]{dsmith@lanl.gov}

\affiliation{Theoretical Division, Los Alamos National Laboratory,
 Los Alamos, NM 87545, USA}


\begin{abstract}
We investigate the role of intra-molecular conformational disorder
and inter-molecular electronic interactions on the electronic
structure of disorder clusters of poly-{\it p}-phenylene vinylene
(PPV) oligomers. Classical molecular dynamics is used to determine
probable molecular geometries, and first-principle density
functional theory (DFT) calculations are used to determine
electronic structure. Intra-molecular and inter-molecular effects
are disentangled by contrasting results for densely packed
oligomer clusters with those for ensembles of isolated oligomers
with the same intra-molecular geometries.  We find that electron
trap states are induced primarily by intra-molecular configuration
disorder, while the hole trap states are generated primarily from
inter-molecular electronic interactions.

\end{abstract}

 \maketitle


The emergence of organic electronic devices, including light
emitting diodes and field effect transistors, fabricated from
conjugated polymers such as poly-{\it p}-phenylene vinylene (PPV)
\cite{Burr90} has stimulated research into the electrical and
optical properties of these semiconducting polymers.  A
theoretical description of these electronic materials is
challenging because they are highly disordered and strong
interactions between molecules occurring in the condensed phase of
the materials are very important in determining their properties.
Despite extensive experimental \cite{Exp} and
theoretical research \cite{theory}, there is still a limited
understanding of these strongly interacting disordered materials.
In this Letter, we use a combination of classical molecular
dynamics (MD) and density functional theory (DFT) to investigate
the electronic structure of densely packed clusters of PPV
oligomers.  Classical MD is used to determine statistically
probable molecular geometries for large clusters of PPV oligomers,
and DFT calculations are used to determine the electronic
structure of the clusters.  The effect of intra-molecular
configurational disorder and of inter-molecular electronic
interactions on the electronic structure are disentangled by
contrasting the calculated density of states and molecular
orbitals for a densely packed oligomer cluster with that for an
ensemble of isolated oligomers with the same intra-molecular
geometries as the oligomers in the clusters.

Statistically probable geometries of disordered clusters of PPV
oligomers were simulated using classical MD.  To isolate the
effect of configurational disorder, we specifically consider
clusters of five-ring PPV oligomers as studied experimentally in
Ref \cite{Hutten99}. Figure 1 shows the structure of the five-ring
PPV oligomer (Fig. 1a), a crystal \cite{Hutten99} of ordered
oligomers (Fig. 1b), and an example of a disordered oligomer
cluster from the MD simulations (Fig. 1c).  The unit cell used for
the MD simulations and DFT calculations consisted of 12 five-ring
oligomers, a total of 816 atoms.

Single crystals of PPV oligomers have been grown and structurally
characterized \cite{Hutten99}.  Thin films of the polymer consist
of small ordered regions, with a structure similar to that of the
crystalline oligomers, separated by disordered regions. The MD
simulations are meant to describe the disordered regions.
Substituted versions of PPV such as poly [2-methoxy,
5-(2'-ethyl-hexyloxy)-1,4-phenylene vinylene] (MEH-PPV) are
frequently used in organic electronics. In these materials the
substituted side groups, added primarily for processing reasons,
are saturated and do not play a direct role in the electronic
structure of the materials. However, in a condensed phase the
substituted side groups tend to separate the conjugated segments
of different molecules and thus reduce inter-molecular electronic
interactions compared to pristine conjugated polymers.  Thus a
combination of the crystal and an ensemble of disordered oligomer
clusters serves as a model for the electronic structure of
pristine PPV. And a combination of the isolated ordered oligomer
and an ensemble of isolated disordered oligomers serves as a model
for the electronic structure of substituted PPVs.

\begin{figure}
   \includegraphics[width=3in]{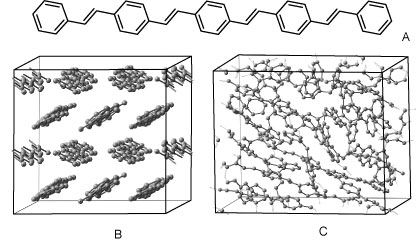}
   \caption{(A) Molecular structure of the ordered oligomer;
(B) Experimental crystal structure \cite{Hutten99}. (C) Molecular
structure of an oligomer cluster determined from the MD
simulation. Hydrogen atoms are not shown for clarity.
\label{fig1}}
\end{figure}

A statistical sample of molecular geometries for the disordered
PPV oligomers was generated by simulated annealing using
periodic-boundary-condition MD with a modified MM3-2000 force
field  \cite{MM3}, as implemented in the Tinker code
\cite{TINKER}. Because the conventional MM3-2000 force field does not
distinguish between single and double C-C bonds, a new term was
added. This new term was chosen to reproduce the C=C double bond
force constant and bond distance obtained from hybrid density
functional calculations (B3LYP). The new force term was used in
linking the vinyl  group and the aromatic $sp^2$ carbon atoms and
the usual MM3-2000 force field was used for the remaining carbons.
This combination of force field terms produced the difference
between averaged length of all single bonds and double bonds
 of 0.19 \AA. This is usually referred as the bond length alteration factor.
 In the simulated annealing
calculations, the PPV oligomers were initially distributed
randomly at extremely low density ($\rho\approx0.01$ $g/cm^3$),
heated to 10,000K, and then pressure was applied to reach the
experimental density of 1.25 $g/cm^3$.
 At this density the volume
of the simulation cell was kept constant and the temperature
cooled to zero K. Molecular geometries were extracted after the
experimental density and zero K were reached, and statistical
samples were generated by repeating the simulated annealing
procedure.

The electronic structure of the crystalline and disordered
oligomer configurations were calculated using
periodic-boundary-condition density functional theory (DFT) at the
generalized gradient approximation (GGA) level of approximation,
using the PW91 functional \cite{PW91}. The DFT
calculations were performed at fixed geometries, determined from
the MD simulations
 using the VASP code \cite{VASP}, VASP with the projector-augmented wave (PAW)
scheme \cite{PAW}.  Due to the large size of the simulation cells
(typically at least 20\AA~ on each side) $\Gamma$-point sampling
of the Brillouin zone was adequate.  This point was explicitly
verified by adding an extra sampling k-point in each direction.
Use of the GGA approximation leads to an underestimate of the
energy gap, typical of pure DFT calculations.  However, we are
primarily interested in electronic states near the HOMO and LUMO
energy levels separately and not in the absolute value of the
energy gap.

\begin{figure}
  \includegraphics[width=3in]{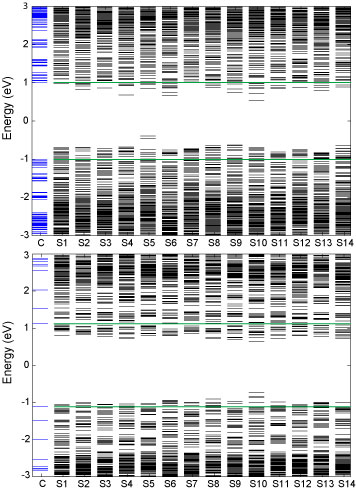}
   \caption{Energy level diagrams from the DFT calculations. The upper
panel shows results for the ideal crystal (far left) and for 14
disordered oligomer clusters whose geometry was determined from
the MD calculations; the lower panel shows results for an isolated
ordered oligomer (far left) and an ensemble of 12 isolated
oligomers with the same molecular geometries as in the
corresponding column of the upper panel. \label{fig2}}
\end{figure}

\begin{figure}
   \includegraphics[width=3in]{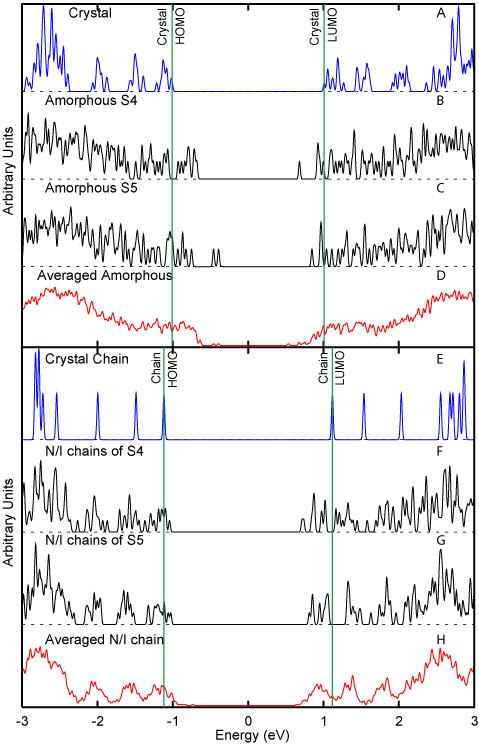}
   \caption{Calculated density of states for: (A) the ideal crystal; (B)
   disordered cluster S4; (C) disordered cluster S5; (D)
   the ensemble average of clusters S1-S14; (E) the ordered
   oligomer in the crystal; (E) the 12 isolated oligomers
   in cluster S4; (F) the 12 isolated oligomers in cluster S5; (H)
   the ensemble average of isolated oligomers in clusters S1-S14. Note
   that 'green lines' denote different energy gaps in top panels A-D vs. bottom
   panels E-H.
    \label{fig3}}
\end{figure}

The upper panel of Fig. 2 shows calculated energy levels for the
ideal crystal (far left) and for 14 different disordered oligomer
clusters, labelled S1-S14, whose geometries were determined from
the MD simulations.  The lower panel of Fig. 2 shows calculated
energy levels for the ensemble of 12 isolated oligomers making up
the oligomer clusters in the corresponding column of the upper
panel. The zero of energy was set to the middle of the energy gap of the
perfect crystal (upper panel) or of the undistorted oligomer
(lower panel). The effect of disorder on the electronic structure of the densely
packed oligomer clusters can be separated into two components:
intra-molecular disorder and inter-molecular electronic
interactions.  Intra-molecular disorder is due to conformational
deformations, resulting from the oligomers not being straight and
planar but, kinked and twisted.  These geometrical distortions
interrupt the conjugation of the $\pi$ orbitals causing the
electronic states to be more localized. Inter-molecular
interactions vary in the densely packed oligomer clusters because
of differences in the the local packing of the oligomers.  The
inter-molecular $\pi$-$\pi$ interactions depend strongly on the
relative orientation of neighboring oligomers.  To disentangle the
effects of intra-molecular conformational disorder and the
inter-molecular interaction effects on the energy level diagram,
the electronic structure of each isolated oligomer in a cluster
was computed independently  (see lower panel of Fig. 2). The lack
of repulsion among electron densities stabilizes the occupied
levels showing that the hole traps originate from the packing of
the oligomers.

Panels A-D of Fig. 3 show calculated density of states for: (A)
the ideal crystal, (B) oligomer clusters S4, and (C) S5, and (D)
the ensemble averaged density of states for oligomer clusters S1 -
S14.  Panels E-H of Fig. 3 show calculated density of states for:
(E) the undistorted oligomer in the ideal crystal, (F) the 12
isolated oligomers in clusters S4, and (G) S5, and (H) the
ensemble averaged density of states for the isolated oligomers in
clusters S1 - S14.  The density of states have been broadened
using a Gaussian with full width at half maximum of 0.0136 eV.
 The zero of energy was set to the middle of the energy
gap of the perfect crystal (panels A-D) or of the undistorted
oligomer (panels E-H).

The density of states for the isolated undistorted oligomer (see,
Fig. 3E) consists of a series of discrete states.  In the crystal
(see Fig. 3A), these discrete states are broadened into bands
because of inter-molecular electronic interactions.  Because of
these inter-molecular interactions, the energy gap of the crystal
is smaller than that of the isolated undistorted oligomer.  The
packing pattern in the crystal is of the herringbone type, a
common arrangement for conjugated oligomers without substitutions
\cite{herringbone}.  Inter-molecular electronic interactions in
crystals with the herringbone structure are relatively small due
to the weak $\pi$-$\pi$ overlap between oligomers  (see Fig. 1B). 
In the ensemble of distorted but
isolated oligomers (see Fig. 3H), the discrete lines of  the
undistorted oligomer are broadened by the statistical distribution
of molecular distortions.  Comparing Fig. 3E with Fig. 3H shows
that this broadening in the valence states  is nearly symmetric
about the position of the corresponding valence state in the
undistorted oligomer.  In contrast, the distortions both broaden
and shift to lower energy the conduction states compared to the
corresponding conduction state in the undistorted oligomer.  This
can be traced to specific intra-molecular interactions. The
occurrence of cis-configurations about the vinylene linkages in
some of the distorted oligomers tends to symmetrically reduce the
HOMO-LUMO energy gap, whereas distortions of the bond lengths and
angles in the vinylene linkages in some distorted oligomers
(either cis or trans configurations) tend to lower both the HOMO
and LUMO energies.  A combination of these two effects leads to
the nearly symmetric broadening of the HOMO levels and the
broadening and shift to lower energy of the unoccupied levels seen
in Fig. 3H. Contrasting Fig. 3A and Fig. 3D, we see stronger
inter-molecular electronic interactions in the disordered oligomer
clusters than in the crystal.  This occurs because the $\pi$-$\pi$
interactions are comparatively weak in the herringbone crystal
structure where adjacent oligomer planes do not strongly overlap,
but can be much larger in the disordered phase where adjacent
oligomer planes can overlap strongly.  Comparing the ensemble
averaged results in Fig. 3D and Fig. 3H (and the corresponding
single cluster cases (panel B with F; and panel C with G)) shows
that inter-molecular interactions broaden and push both valence
and conduction states to higher energy.

High energy valence states can act as hole traps and low energy
 conduction states can act as electron traps. As seen in
Fig. 3, intra-molecular distortions, on average, symmetrically
broaden valence states and broaden and push conduction states to
lower energy, whereas inter-molecular electronic interactions, on
average, broaden and push both valence and conduction states to
higher energy.  As a result, electron trap states are favored by
intra-molecular configurational disorder, and hole trap states are
favored by inter-molecular electronic interactions.  In
substituted polymers, such as MEH-PPV, inter-molecular electronic
interactions are reduced compared to pristine polymers. Thus, hole
traps should be more dominant in pristine PPV and electron traps
more dominant in PPVs with long side-chains.

\begin{figure}
   \includegraphics[width=3in]{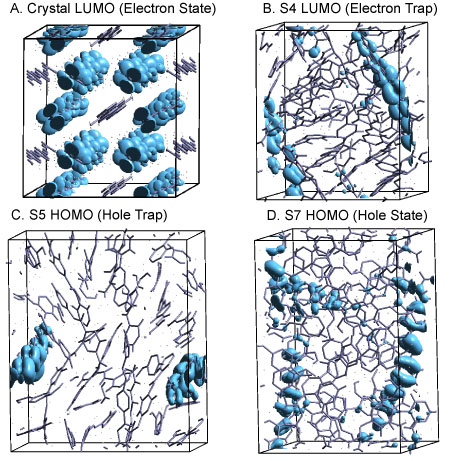}
   \caption{Calculated electron densities for selected states: (A) the crystal
   LUMO state; (B) the LUMO level for cluster S4; (C) the HOMO level for cluster S5; and
   (D) the HOMO level for cluster S7.
 \label{fig4}}
\end{figure}

Fig. 4 shows the electron densities (squared molecular orbital
amplitude) for selected one electron states.  Fig. 4A shows the
electron density for the LUMO of the oligomer crystal. This state,
like all the other states in the crystal, is delocalized
throughout the crystal.  By contrast, the HOMO and LUMO states in
the disordered oligomer clusters are generally localized on
individual molecules, whereas states further from the energy gap
are often delocalized on more than one oligomer.  Fig. 4B shows
the LUMO level for cluster S4.  This state (see Fig. 2) is
separated in energy from the other states in the cluster and lies
deep within the energy gap of the corresponding crystal.  It is
strongly localized on a single oligomer.  Fig. 4C shows the HOMO
level for cluster S5.  This state (see Fig. 2) is also separated
in energy from the other states in the cluster, in the energy gap
of the corresponding crystal, and strongly localized.  Fig. 4D
shows the HOMO level for cluster S7.  This state (see Fig. 2) is
not strongly separated in energy from the other states in the
cluster and not as deep into the energy gap of the corresponding
crystal as the LUMO level on cluster S4 or the HOMO level on
cluster S5. It is delocalized on several oligomers.  The deep trap
levels in the disordered clusters are usually localized on a
single molecule.  The S4 LUMO is an example of an electron trap
localized on one oligomer, and the HOMO of S5 is an example of a
hole trap localized on one oligomer.

In summary, we used a combination of classical molecular dynamics
and density functional theory to investigate the role of
intra-molecular conformational disorder and inter-molecular
electronic interactions  on the electronic structure of disordered
clusters of poly-{\it p}-phenylene vinylene (PPV).  We found that
electron trap states are induced primarily by intra-molecular
configurational disorder, while the hole trap states are generated
primarily from inter-molecular electronic interactions.  Thin
films of PPV consist of small ordered regions, with a local
structure similar to that of the crystalline oligomers, separated
by disordered regions.  In substituted PPVs such as MEH-PPV the
side groups separate the conjugated segments of different
molecules and reduce inter-molecular electronic interactions
compared to pristine PPV. Our model suggests that hole traps
should be more dominant in pristine PPV, compared to PPV with long
side-chains due to the weak  inter-molecular electronic interactions 
in the latter. Because traps are important in
determining electrical transport properties, these results provide
strategies to design materials for good electrical transport.

This work was supported by DOE Office of Basic Energy Sciences
under Work Proposal Number 08SCPE973. LANL is operated by Los Alamos
National Security, LLC, for the National Nuclear Security Administration
of the U.S. Department of Energy under contract DE-AC52-06NA25396.


\end{document}